\documentclass[11pt,a4paper,showpacs,aps]{revtex4-1}
\usepackage[latin1]{inputenc}
\usepackage{amssymb}
\usepackage{amsmath}
\usepackage{graphicx}
\usepackage{fancyhdr}

\begin{document}

\title{D=(2+1) O(N) Wess-Zumino model in a large N limit}

\author{A.~C.~Lehum}
\email{andrelehum@ect.ufrn.br}
\affiliation{Escola de Ci\^encias e Tecnologia, Universidade Federal do Rio Grande do Norte\\
Caixa Postal 1524, 59072-970, Natal, RN, Brazil}

\begin{abstract}
Using the superfield formalism, the effective K\"{a}hlerian superpotential of the massless ${\cal{N}}=1~O(N)$  Wess-Zumino model is computed in the limit of large $N$, in three spacetime dimensions. The effective K\"{a}hlerian superpotential is evaluated at the subleading order in the $1/N$ expansion, which involves diagrams up to two-loop order, for a small coupling constant. We show that the $O(N)$ symmetry of the model is preserved in this approximation and that no mass is dynamically generated in the supersymmetric phase. We discuss why spontaneous $O(N)$ symmetry breaking cannot be induced by radiative corrections in such model.      
\end{abstract}


%\keywords{Wess-Zumino model, spontaneous symmetry breaking, effective superpotential}

\maketitle

\section{Introduction}

Supersymmetry (SUSY) is one of most important proposals of physics beyond the Standard Model. It appeared for the first time in two different scenarios. It was discovered as a type of gauge symmetry of the string when fermionic states are present, and in another scenario, it was proposed as a way to avoid the Coleman-Mandula theorem, i.e., an extension of the spacetime algebra was found, namely, the super-Poincaré algebra. Then, the first supersymmetric action as a four-dimensional field theory was proposed by Wess and Zumino in 1974~\cite{Wess:1974tw}. Although the so-called Wess-Zumino model is not a realistic theory to describe the physics beyond the Standard Model, it has played an important role in studying several aspects of supersymmetric theories. Indeed, toy models are widely used as theoretical laboratories because they can exhibit the same wealth of more realistic theories, and often give an intuition of nature's behaviour.

In this context, studying effective potentials is an important tool to understand, through classical concepts, the quantum behaviour of physical systems. Moreover, effective potentials are a natural way to study spontaneous symmetry breaking. Of special interest, the $1/N$ expansion~\cite{'tHooft:1973jz} has been a significant instrument responsible for remarkable results, e.g., the spontaneous breaking of scale invariance in three-dimensional supersymmetric models~\cite{Bardeen:1984dx}, studies of the renormalization group through finite temperature quantum field theory techniques~\cite{Synatschke:2010ub}, and the equivalence between $(\phi^2)^2$ and non-linear sigma models in the description of critical phenomena~\cite{Brezin:1975sq,Biscari:1990ty}. The large $N$ technique is also applied in the study of the relation between Abelian Higgs and $CP(N-1)$ models~\cite{D'Adda:1978kp}. For further information about the $1/N$ expansion in Quantum Field Theory, as well as modern applications, see \cite{Moshe:2003xn}.

Recent interest in three-dimensional supersymmetric theories comes from the fact that a wide class of models could be related to M2-branes~\cite{Bagger:2006sk,Krishnan:2008zm,Gustavsson:2007vu}. In such models, superconformal invariance is an important ingredient. On the other hand, spontaneous breaking of conformal symmetry induced by radiative corrections~\cite{Coleman:1973jx} was shown to be a possible effect in some three-dimensional models~\cite{Dias:2003pw,Lehum:2008vn,Ferrari:2010ex}. The three-dimensional massless Wess-Zumino model with a quartic superfield self-interaction exhibits superconformal invariance, and furthermore it can be interpreted as the matter sector of the SUSY Chern-Simons-matter models~\cite{csmm}.

In the present paper, the effective K\"{a}hlerian superpotential~\cite{BK0} of the massless ${\cal{N}}=1$ $O(N)$ Wess-Zumino model is evaluated at subleading order in the large $N$ expansion in three-dimensional spacetime, showing that no generation of mass is induced by Coleman-Weinberg mechanism~\cite{Coleman:1973jx}. This is developed using the superfield formalism, therefore the discussion is restricted to the supersymmetric phase. This Brief Report intends to discuss some aspects of the $O(N)$ Wess-Zumino model ground state that was not approached in~\cite{Ferrari:2009zx}.

\section{${\cal{N}}=1~O(N)$ Wess-Zumino model}

The action of the ${\cal{N}}=1$ O(N) Wess-Zumino model in the $D=(2+1)$ superspace can be defined as
\begin{eqnarray}\label{eq01}
S&=& \int{d^3x~d^2\theta}\Big{\{}\frac{1}{2}\Phi_a D^2\Phi_a +\frac{g}{4}(\Phi_a\Phi_a)^2\Big{\}},
\end{eqnarray}

\noindent where $a=1,2,\cdots,N$. The conventions and notations are adopted as in~\cite{Gates:1983nr}. The real superfields $\Phi_a$ are expanded in a Taylor series in the Grassmaniann variable as
\begin{eqnarray}
\Phi_a(x,\theta)=\phi_a(x)+\theta^{\alpha}\psi_{a\alpha}(x)-\theta^2F_a(x),
\end{eqnarray}

\noindent where $\phi$ and $F$ are real scalar fields and $\psi$ is a two component Majorana fermion.

We will evaluate the effective K\"{a}hlerian superpotential in the large $N$ limit, using some of the techniques described in~\cite{Ferrari:2009zx}. To do this, let us rescale $g\rightarrow\lambda/N$ and let $\langle\Phi_N\rangle=\sqrt{N}\varphi$, where $\varphi$ is a constant classical (background) superfield given by $\varphi=\varphi_1-\theta^2\varphi_2$. The rescaled coupling stands for the proper expansion parameter in the theory, where the large $N$ limit is accomplished by taking $N\gg1$, with the 't Hooft parameter $\lambda$ being fixed. After these rescalings, the action (\ref{eq01}) can be rewritten as  
\begin{eqnarray}\label{eq2}
S&=&\int{d^5z}\Big{\{}\frac{1}{2}\Phi_j (D^2+\lambda\varphi^2)\Phi_j
+\frac{1}{2}\Phi_N (D^2+3\lambda\varphi^2)\Phi_N 
+\frac{\lambda}{4N}(\Phi_j^2+\Phi_N^2)^2 \nonumber\\
&&+\frac{\lambda}{\sqrt{N}}\varphi\Phi_N(\Phi_j^2+\Phi_N^2)
+\sqrt{N}(D^2\varphi+\lambda\varphi^3)\Phi_N+\frac{N\lambda}{4}\varphi^4\Big{\}}.
\end{eqnarray}

In general, an effective superpotential in a three-dimensional spacetime has the form
\begin{eqnarray}\label{eq3}
V_{eff}(\varphi)&=&{\mathcal F}(D^{\alpha}\varphi D_{\alpha}\varphi,\,D^2\varphi,\,\varphi)+{\mathcal K}(\varphi),
\end{eqnarray}

\noindent where ${\mathcal F}(D^{\alpha}\varphi D_{\alpha}\varphi,\,D^2\varphi,\,\varphi)$ is a superpotential where some supercovariant derivative $D^{\alpha}$ appears applied to the background superfield $\varphi$, and ${\mathcal K}(\varphi)$ is the K\"{a}hlerian superpotential characterized by the absence of supercovariant derivatives $D^{\alpha}$. 

Integrating (\ref{eq3}) in $d^2\theta$ we get: 
\begin{eqnarray}\label{+1}
U_{eff}=\!\!\int{d^2\theta}V_{eff}(\varphi)=\!\!\int{d^2\theta}\left[{\mathcal F}(D^{\alpha}\varphi D_{\alpha}\varphi,\,D^2\varphi,\,\varphi)+{\mathcal K}(\varphi)\right]=\varphi_2\frac{d{\mathcal K}}{d\varphi_1}(\varphi_1)+\varphi_2^2 f(\varphi_1,\varphi_2),
\end{eqnarray}

\noindent where in the last term we have to note that the contributions coming from the $F$-term start with, at least, two powers of $\varphi_2$.

The conditions that minimize $U_{eff}$ are given by:
\begin{eqnarray}
&&\frac{\partial U_{eff}}{\partial \varphi_1}=\varphi_2 \frac{d^2{\mathcal K}}{d \varphi_1^2}(\varphi_1)+ 
\varphi_2^2\frac{\partial f}{\partial \varphi_1}(\varphi_1,\varphi_2)=0,\\
&&\frac{\partial U_{eff}}{\partial \varphi_2}=\frac{d {\mathcal K}}{d\varphi_1}(\varphi_1)
+2 \varphi_2 f(\varphi_1,\varphi_2)+\varphi_2^2 \frac{\partial f}{\partial \varphi_2}(\varphi_2,\varphi_2)=0.
\end{eqnarray}

\noindent For $\varphi_2=0$, suggesting the supersymmetric phase, these equations imply that the minimum of $U_{eff}$ is zero if only if
\begin{eqnarray}\label{condmin}
\frac{d{\mathcal K}}{d\varphi_1}(\varphi_1)=0.
\end{eqnarray}

If there exists a solution $\varphi_1=v$ of the above equation, we have that $U_{eff}(v,0)=0$, preserving supersymmetry. Therefore, the knowledge of the K\"{a}hlerian superpotential is enough to decide on the possibility of spontaneous supersymmetry breaking~\cite{Ferrari:2010ex,Burgess:1983nu}.

The 't Hooft coupling $\lambda$ is fixed, and to simplify our analysis let us consider it small, i.e., $\lambda\ll1$. This approximation allows us to truncate the series of Feynman diagrams that contribute to the K\"{a}hlerian effective superpotential. 

\begin{figure}[ht] \begin{center} \includegraphics[height=4cm ,angle=0 ,width=6cm]{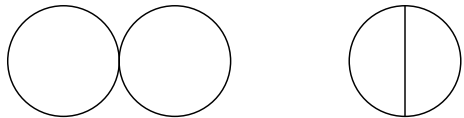} \end{center} \caption{\em Topologies of two-loop diagrams that contribute to the effective superpotential.} \label{2loop} \end{figure}

We next evaluate the effective superpotential up to order $\lambda^3$ at subleading order in the $1/N$ expansion, corresponding to including up to the two-loop diagrams, whose topologies are drawn in Fig. \ref{2loop}. The tree level contribution is easily identified from (\ref{eq2}) as
\begin{eqnarray}\label{eq4}
{\mathcal K}(\varphi)=-N\frac{\lambda}{4}\varphi^4~,
\end{eqnarray}

\noindent and the one-loop contributions are given by the trace of the determinants. These traces can be evaluated as described in~\cite{Ferrari:2009zx}. Then, the one-loop contribution to the effective action can be written as
\begin{eqnarray}\label{eq5}
\Gamma^{(1)}&=&\frac{i}{2}\mathrm{Tr}\ln\left[\Box-{\mathcal K}''(\varphi)D^2\right]=\frac{1}{16\pi}\int{d^5z}(N+8)\lambda^2\varphi^4~,
\end{eqnarray}

\noindent where the regularization by dimensional reduction~\cite{Siegel:1979wq} is used to perform the integrals.

The two-loop diagrams, Fig. \ref{2loop}, contribute with
\begin{eqnarray}\label{eq6}
\Gamma^{(2)}&=&i\int{d^5z}\int{\frac{d^3k}{(2\pi)^3}\frac{d^3q}{(2\pi)^3}}\Big{\{}\frac{5\lambda^3\varphi^4}{(k^2+M^2)(q^2+M^2)[(k+q)^2+m^2]}\nonumber\\
&&-N\frac{\lambda}{4}\frac{1}{(k^2+M^2)(q^2+M^2)}
-N\frac{\lambda}{2}\frac{1}{(k^2+M^2)(q^2+m^2)}
\Big{\}}.
\end{eqnarray}

Performing the integrals using the formulae given in~\cite{Dias:2003pw,Tan:1996kz} and adding all contributions, the effective action can be written as
\begin{eqnarray}\label{eq7}
\Gamma&=&i\int{d^5z}\Big{\{}-N\frac{\lambda}{4}\varphi^4+(N+8)\frac{\lambda^2}{16\pi}\varphi^4
+N\frac{7\lambda^3}{64\pi^2}\varphi^4+\frac{5\lambda^3}{16\pi^2}\varphi^4\left[C(\epsilon)-\ln{\frac{\varphi^2}{\mu}}\right]+\mathcal{L}_{CT}\Big{\}},
\end{eqnarray}

\noindent where $\mathcal{L}_{CT}= NC\varphi^2+N\lambda B \varphi^4$ is the Lagrangian of counterterms, $\mu$ is a mass scale introduced by the regularization, $C(\epsilon)=+\dfrac{1}{2}(\dfrac{1}{\epsilon}-\gamma+1)+\ln{\dfrac{2\sqrt{\pi}}{5\lambda}}$, $\gamma$ is the Euler-Mascheroni constant and $\epsilon=(3-D)$, with $D$ being the number of spacetime dimensions. We observe the presence of a divergent term, which requires a renormalization condition to remove it. 

Just as in the original proposal of Coleman and Weinberg~\cite{Coleman:1973jx}, the renormalization conditions adopted here are
\begin{subequations}
	\begin{equation}\label{eq8a} 
	\frac{\partial^2{\mathcal K}(\varphi)}{\partial\varphi^2}\Big{|}_{\varphi=0}=0~,
	\end{equation}
	\begin{equation}\label{eq8b} 
	\frac{\partial^4{\mathcal K}(\varphi)}{\partial\varphi^4}\Big{|}_{\varphi=v}=-4!\frac{\lambda}{4}N~,
	\end{equation}
\end{subequations}

\noindent where $v$ is the renormalization scale. In four dimensions, a quadratic divergence appears at one-loop order, and the mass counterterm $C$ is used to remove it. In three dimensions, the regularization by dimensional reduction avoids any divergence at the one-loop level, hence no mass renormalization is necessary to ensure the renormalizability of the model, consequently the renormalization condition (\ref{eq8a}) implies $C=0$.

The condition (\ref{eq8b}) determines the counterterm $B$ as
\begin{eqnarray}\label{eq8c}
B=-\frac{\lambda}{192N\pi^2}\left(96\pi+12N\pi+250\lambda+21N\lambda-60\lambda c(\epsilon)+60\lambda\ln{\frac{v^2}{\mu}} \right).
\end{eqnarray}

Substituting the expression for $B$ into (\ref{eq7}), the renormalized K\"{a}hlerian effective superpotential can be written as 
\begin{eqnarray}\label{eq9}
{\mathcal K}(\varphi)&=&-N\frac{\lambda}{4}\varphi^4+\frac{5}{16\pi^2}\lambda^3\varphi^4\left(\frac{25}{6}-\ln{\frac{\varphi^2}{v^2}}\right).
\end{eqnarray}

The condition that minimizes the effective superpotential is
\begin{eqnarray}\label{eq10}
\frac{\partial {\mathcal K}(\varphi)}{\partial\varphi}&=&0~,
\end{eqnarray}

\noindent which has three solutions: 
\begin{subequations}
	\begin{equation}\label{eq11a} 
	\varphi=0~,
	\end{equation}
	\begin{equation}\label{eq11b} 
	\varphi=\pm v~\mathrm{exp}\left(\frac{11}{6}-\frac{4\pi^2N}{10\lambda^2}\right)~.
	\end{equation}
\end{subequations}

The first one is the trivial solution and no mass is generated by the radiative corrections. The second and third ones, at first glance, would generate mass to the scalar superfields. But, to this solution, it is expected that the minimum of the effective potential lies around $\varphi=v$. This is satisfied when the exponential function of $\lambda$  is approximately $1$. So the exponent should satisfy 
\begin{eqnarray}\label{eq12}
\frac{11}{6}-\frac{4\pi^2N}{10\lambda^2}\approx 0,
\end{eqnarray}

\noindent which fixes $\lambda$ to be of the order of $\sqrt{N}$. Once it is admitted that $N\gg 1$ this result contradicts the initial condition that $\lambda$ should be much less than $1$, invalidating the perturbative expansion. 

In general, this issue seems to be a characteristic of the Coleman-Weinberg mechanism when the model has only one coupling constant. The condition that constraints $\lambda$ be large is improved when the global symmetry of the model is promoted to a gauge one. In a gauge theory situation, what appears is a condition that constrains the self-interaction coupling constant ($\lambda$) to be of the order of some power of the gauge coupling constant, where this power depends on the number of spacetime dimensions. Therefore, it is natural to expect that a gauge version of the model presented here can exhibit a consistent generation of mass through the Coleman-Weinberg mechanism~\cite{Ferrari:2010ex}.

Taking the limit of (\ref{eq11b}) when $N$ tends to infinity, we see that $\varphi$ goes to zero. The fact is that the only consistent solution to the minimum of the effective K\"{a}hlerian superpotential is $\varphi=0$, indicating that in the supersymmetric phase no generation of mass is possible in this model. 

\section{Final remarks}

In summary, in this work the effective K\"{a}hlerian superpotential of the massless $O(N)$ Wess-Zumino model was computed at subleading order in the large N limit in three-dimensional spacetime. The effective K\"{a}hlerian superpotential was evaluated keeping the 't Hooft coupling small. This approximation allowed us to truncate the series of Feynman diagrams at the two-loop corrections. This choice is justified considering that a mass scale is always introduced in this model through logarithmic divergent diagrams, whose first appearance is at the two-loop Feynman graphs. 

This evaluation of the effective K\"{a}hlerian superpotential allow us to affirm that supersymmetry can not be spontaneously broken at the considered approximation, because condition (\ref{eq10}) is satisfied.  Furthermore, no generation of mass is induced by radiative corrections in the approximation presented here. The approximation adopted here relies on two distinct approximation approaches, expansion in powers of $1/N$ and in powers of $\lambda$. All of our conclusions is limited to large $N$ and small $\lambda$. A similar procedure was adopted in the study of $1/N$ expansion of an $U(N)$ gauge model~\cite{Kang:1976ss}. Only a full $1/N$ evaluation, i.e., without restrictions over $\lambda$, can clarify about the spontaneous generation of mass in such model, being a question that we have not succeeded in answering in this paper. But, we have found that if $O(N)$ Wess-Zumino model can present dynamical generation of mass, such effect should be a non-perturbative phenomenon.

As a final remark, we comment about the possibility of noncommutative extensions of the present article. Lately, noncommutative extensions of the ordinary field theories have been intensely discussed in the literature because such theories are related with certain low energy limits of string theory~\cite{Seiberg:1999vs}. In particular, a four-dimensional Wess-Zumino model was shown to be a consistent noncommutative field theory, free of the dangerous ultraviolet/infrared mixing to all orders in perturbation theory~\cite{Girotti:2000gc}. A study of the dynamical generation of mass in a three-dimensional version of such model is in progress.

\vspace{.5cm}
{\bf Acknowledgments.} This work was partially supported by the Brazilian agency Conselho Nacional de Desenvolvimento Cient\'{\i}fico e Tecnol\'{o}gico (CNPq).

\vspace{2cm}

\end{document}